\documentclass[prb,aps,floats,showpacs,amsmath,amssymb,preprint]{revtex4}
\usepackage{bm,graphicx,epsf}
\newcommand{\be}{\begin{equation}}
\newcommand{\ee}{\end{equation}}   
\newcommand{\bea}{\begin{eqnarray}}
\newcommand{\eea}{\end{eqnarray}}
\newcommand{\phrl}[1]{Phys.~Rev.~Lett. {\bf #1}}
\newcommand{\phrb}[1]{Phys.~Rev.~B {\bf #1}}

\newcommand{\bib}{\bibitem}
\newcommand{\lb}{\left[}
\newcommand{\rb}{\right]}
\newcommand{\lp}{\left(}
\newcommand{\rp}{\right)}

\renewcommand{\k}{{\bf k}}

\begin{document}

\title{Manifestation of helical edge states as zero-bias magneto-tunneling-conductance peaks in non-centrosymmetric superconductors}
\author{Soumya P. Mukherjee and Sudhansu S. Mandal}
\affiliation{Department of Theoretical Physics, Indian Association for the Cultivation of Science, Jadavpur, Kolkata 700 032, India}

\date{\today}


\begin{abstract}
 Helical edge states exist in the mixed spin-singlet and -triplet phase of a noncentrosymmetric
superconductor (NCSS) 
when the pair amplitude (PA) in the negative helicity band, $\Delta_-$, is smaller than the PA in the 
positive helicity band, $\Delta_+$, i.e., when
the PA in the triplet component is more than the same in the singlet component.
We numerically determine energies of these edge states as a function of $\gamma = \Delta_-/\Delta_+$.   
The presence of these edge states is reflected in the tunneling process from a normal metal to an NCSS
across a bias energy $eV$: (i) Angle resolved spin conductance (SC) obeying the symmetry $g_s(\phi)
=-g_s(-\phi)$ shows peaks when the bias energy equals the available quasiparticle edge state energy provided
$\vert eV \vert \lesssim \Delta_-$. (ii) The total SC, $G_s$, is zero but modulates with $eV$
for finite magnetic field $H$. (iii) The zero bias peaks of $G_s$ and total charge conductance, $G_c$, at finite 
$H$ split into two at finite $eV$ for moderate $H$. (iv) At zero bias, $G_c$ and $G_s$ increase with $H$ and show
peaks at $\vert H\vert \sim \gamma H_0$ where $H_0$ is a characteristic field.
\end{abstract}

\maketitle

\section{Introduction}

Recently discovered noncentrosymmetric (NCS) superconductors such as
CePt$_3$Si (Ref.\onlinecite{Bauer}) and Li$_2$Pt$_3$B (Ref.\onlinecite{Yuan}) having strong spin-orbit 
interaction (SOI) among various types of unconventional superconductors are of current interest in their own right. 
Besides, superconductivity at the interfaces, such as the interface between LaAlO$_3$/SrTiO$_3$
(Ref.\onlinecite{Reyren}), may also be classified as the two-dimensional NCS superconductivity due to the strong potential gradient. 
The SOI in NCS superconductors
leads to mixture of the spin- singlet ($s$-wave) and triplet ($p$-wave) pairing \cite{Sigrist}; the pairing amplitude
in positive (negative) helicity band is $\Delta_+$ ($\Delta_-$) with $p_y - ip_x$ symmetry.  
The triplet pairing occurs in both $s_z=-1$ and $+1$ channels but their chiral $p$-wave symmetries are conjugate \cite{Mandal}
to each other,  
where $s_z$ is the the spin component of a triplet pair along quantization direction.
Thus nonmagnetic NCS superconductors have potential of producing spin current without magnetic field.
These may produce Josephson spin tunneling current \cite{Mandal} between two NCS superconductors
and spin tunneling current \cite{Nagaosa} due to Andreev reflection \cite{Andreev} across the junction between a normal metal
and NCS superconductor.
Both the up- and the down-spin holes will be reflected in the Andreev process; consequently the
spin polarized tunneling current flows.

There exists helical edge mode \cite{Vekhter,Nagaosa} when the superconductor has 
more triplet component with $p_y \pm ip_x$ symmetry than singlet component. 
The low energy Andreev reflection is
mostly due to these edge modes and the incident angle dependent spin polarized current flows \cite{Nagaosa}
through the interface. In
the presence of magnetic field, the incident-angle-integrated current is also spin polarized. There is no helical edge
mode for purely $s$-wave symmetry. 
The existence of zero energy Majorana fermions at the vortex state and their
obeying non-abelian statistics \cite{Read} is also a possibility in the NCS superconductors \cite{Fujimoto} {\em a la}
chiral $p$-wave superconductor \cite{Stone} such as Sr$_2$RuO$_4$ (Ref.\onlinecite{Maeno}).

The helical edge state is present \cite{Vekhter,Nagaosa} in the NCS superconductors when the magnitude of the triplet
component of the pair amplitude is larger than the singlet component, i.e., when the ratio
between pair amplitudes in negative and positive helicity bands, $\gamma = \Delta_-/\Delta_+ > 0\,\, , \,\, (\Delta_- < \Delta_+)$.
Applying boundary condition at the edges, Tanaka {\em et al} \cite{Nagaosa} have found that 
the bound state energy $E$ is proportional to transverse momentum $k_y$ for small $k_y$. 
In this article, we numerically obtain
the energy of the edge states for all permissible $k_y$, since all of these have role in the tunneling process.
We find that the midgap quasiparticle energy $(E<\Delta_-)$ for the edge state decreases with $\gamma$.

 Although the tunneling charge and spin conductances for {\it purely} triplet symmetry (i.e., for $\gamma =1$) 
have been studied by Tanaka {\em et al} \cite{Nagaosa}, 
exploration for the mixed triplet and singlet symmetries is necessary since in the system like 
Li$_2$Pt$_3$B, triplet and singlet components are comparable \cite{Yuan}. We employ the method of
Tanaka {\em et al} \cite{Nagaosa} and study tunneling conductances
for different proportionate mixture of triplet and singlet components ($\gamma \neq 1$) here in this article and
find new and interesting consequences. 
The angle resolved spin current, denoted as $g_s(\phi)$,
 shows peaks at those values of incident angle $\phi$
for which the energy of the incident electron is equal to the quasiparticle bound state energy, provided the bias energy
$\vert eV\vert \lesssim \Delta_-$ and it obeys the symmetry $g_s(\phi) = - g_s(-\phi)$ and hence
total spin conductance $G_s$ is zero at zero magnetic field. However, at finite magnetic field $G_s$ is finite
and obeys the symmetry $G_s(eV,H) = -G_s(-eV,H) = -G_s(eV,-H)$ . The total charge conductance
$G_c$ shows a dip at the bias energy $\vert eV \vert = \Delta_-$, a zero bias peak (ZBP) at zero magnetic field, 
splitting of the peak into two at finite bias and a dip at zero bias for moderate magnetic field,
and then reappearance of the ZBP at higher magnetic field before it eventually vanishes at very high magnetic
field. Although $G_s$ is zero at zero magnetic field, it shows ZBP at finite magnetic field. The splitting of
peaks and the shifting of peaks at finite bias with the increase of magnetic field is similar as in the case of $G_c$.
The zero bias magnitude of both $G_c$ and $G_s$ increases with $\vert H \vert$ and show peaks at $\vert H\vert
\sim \gamma H_0$ with $H_0 = \Phi_0/(\pi^2\xi\lambda_d)$ which is the characteristic field where $\Phi_0$ is the flux
quantum, $\xi$ is the coherence length and $\lambda_d$ is the penetration length of the superconductor.

The article is organized as follows. In Sec.II, we derive an equation for the quasiparticle
energy of the helical edge state in noncentrosymmetric superconductor using the boundary condition of
forming bound states. This equation is
numerically solved to find the energies of the quasiparticle bound states. The tunneling charge
and spin conductances from a normal metal to a NCS superconductor in absence and presence of magnetic field 
are formulated in section III. The conductances are numerically determined and the results are presented in
Sec.IV. We summarize our results in section V.

\section{Helical Edge State}

 We begin with the Hamiltonian for an NCS superconductor in which Cooper pairs form between the electrons 
 within the same spin-split band: 
\be
{\cal H} = \sum_{\k , \lambda = \pm} \lb \xi_{\k\lambda }c_{\k \lambda}^\dagger
c_{\k \lambda} +\lp \Delta_{\k\lambda} 
c_{\k \lambda}^\dagger c_{-\k \lambda}^\dagger + \, \rm{h. c.}
 \rp \rb  \, ,
\label{H_band}
\ee
where $\xi_{\k\lambda} = \xi_{\k}+\lambda \alpha \vert \k\vert$ for Rashba SOI \cite{Rashba}, 
$\xi_{\k} = \hbar^2\k^2/(2m)-\mu$. Here $\mu$, $m$, $\lambda$, $\k$, $\alpha$, and $\Delta_{\k\lambda}$ 
denote chemical potential, mass of an electron, spin-split band index $(\pm)$, momentum of an electron,
coupling constant of Rashba SOI given by $\hat{V}_{so}=\alpha \bm{\eta}_\k \cdot \hat{\bm{\sigma}}$ with
$\bm{\eta}_\k = \hat{\bm{y}}k_x - \hat{\bm{x}}k_y$ and the Pauli matrices $\bm{\sigma}$, 
and pair potential in band $\lambda$ respectively.
We choose $k_y +ik_x$-wave pair in both the bands, i.e., $\Delta_{\k\lambda} = \Delta_\lambda \Lambda_\k$
with $\Lambda_\k = -i\exp [-i\phi_\k]$. This corresponds to triplet component of pair potential
$\hat{\Delta}_T = (\bm{d}_\k \cdot \bm{\sigma})i\sigma_y$ with 
$\bm{d}_\k = \frac{1}{2\vert \k\vert}(\Delta_+ + \Delta_-)\bm{\eta}_\k$, i.e., the amplitude of the
triplet component $\Delta_t= \frac{1}{2}(\Delta_+ +\Delta_-)$ and the singlet component of the pair potential
is $\hat{\Delta}_S = i\Delta_s \sigma_y$ with amplitude $\Delta_s = \frac{1}{2}(\Delta_+ -\Delta_-)$ 
(Ref.\onlinecite{Mandal}). 
Therefore the superconductor is purely triplet with $k_y +i k_x$-wave
symmetry when $\Delta_+ = \Delta_-$, purely singlet with $s$-wave symmetry when $\Delta_- = -\Delta_+$,
and triplet and singlet components with equal amplitude when $\Delta_- = 0$.
Therefore the Hamiltonian (\ref{H_band}) in the matrix form \cite{Sigrist} read as
\be
  H = \left( \begin{array}{cc}
       \hat{h}_\k & \hat{\Delta}_\k \\
      -\hat{\Delta}^*_{-\k} & -\hat{h}^*_{-\k} \end{array} \right) \, ,
\label{H_super}
\ee
where $\hat{h}_\k = \xi_\k + \hat{V}_{so}$ and $\hat{\Delta}_\k = \hat{\Delta}_T +\hat{\Delta}_S$.
The solution of the Hamiltonian (\ref{H_super}) in the bulk 
is with the energy eigen values $\pm \sqrt{\xi_+^2 + \Delta_+^2}$ and $\pm \sqrt{\xi_-^2 + \Delta_-^2}$,
in  with the Cooper pairing between electrons within the same spin-split band.
Correspondingly, there are two Fermi surfaces with Fermi momenta $k_F^\pm = \mp m\alpha /\hbar^2 +
\sqrt{(m\alpha/\hbar^2)^2 + 2m\mu/\hbar^2}$, i.e., $k_F^+ < k_F^-$.

  Consider a two-dimensional semi-infinite NCS superconductor
 with the edge along $y$-direction such that the edge is located at $x=0$ and the superconductor is in the region $x>0$.
 We then mix two quasiparticle and two quasihole states at and near the edge.
The corresponding wave function will have the form as
\bea
\Psi_S (x,y) &=& e^{ik_yy} [ e^{-\kappa_+x} \{ c_1\, \psi_e^+ e^{ik^+_{Fx}x} + c_2\, \psi_h^+ e^{-ik^+_{Fx}x} \}
  \nonumber\\
   & & + e^{-\kappa_-x} \{ d_1 \,\psi_e^- e^{ik^-_{Fx}x} + d_2\, \psi_h^- e^{-ik^-_{Fx}x} \} ] \, ,
\eea
where Fermi momenta along $x$-direction in two spin-split bands are $k_{Fx}^\pm = \sqrt{k_F^{\pm 2}-k_y^2}$.
Quasiparticle and quasihole wave functions \cite{Nagaosa} in two spin-split bands $(\pm)$ are given by
\bea
 & &\psi_e^+ = \left( \begin{array}{c} u_+ \\  - ie^{i\phi_+}u_+ \\  ie^{i\phi_+}v_+ \\ v_+ \end{array} \right) \, ,\,
 \psi_h^+ = \left( \begin{array}{c} v_+ \\  + ie^{-i\phi_+}v_+ \\ - ie^{-i\phi_+}u_+ \\ u_+ \end{array} \right) \, ,
\label{psi_plus} \\
 & &\psi_e^- = \left( \begin{array}{c} u_- \\   ie^{i\phi_-}u_- \\  ie^{i\phi_-}v_- \\ -v_- \end{array} \right) \, ,\,
 \psi_h^- = \left( \begin{array}{c} v_- \\  - ie^{-i\phi_-}v_- \\  -ie^{-i\phi_-}u_- \\ -u_-\end{array} \right) \, ,
\label{psi_minus}
\eea
with  $\frac{u_+}{v_+} = (E-i\Gamma_+)/\Delta_+$, $\frac{u_-}{v_-} = (E-i\Gamma_-)/\Delta_-$, and 
$\Gamma_\pm = \sqrt{\Delta_\pm^2-E^2}$
 for an edge state with
energy $E$, and $\sin (\phi_\pm) = k_y/k_F^\pm$. 
Here $c_1$, $c_2$, $d_1$, and $d_2$ are the corresponding weights at which these four quasiparticle and
quasihole states mix, and $\kappa_\pm = m\Gamma_\pm /k_{Fx}^\pm$
 are the inverse of the length scales of localized edge state for two spin-split bands.

\begin{figure}
\centerline{\epsfysize=8cm\epsfbox{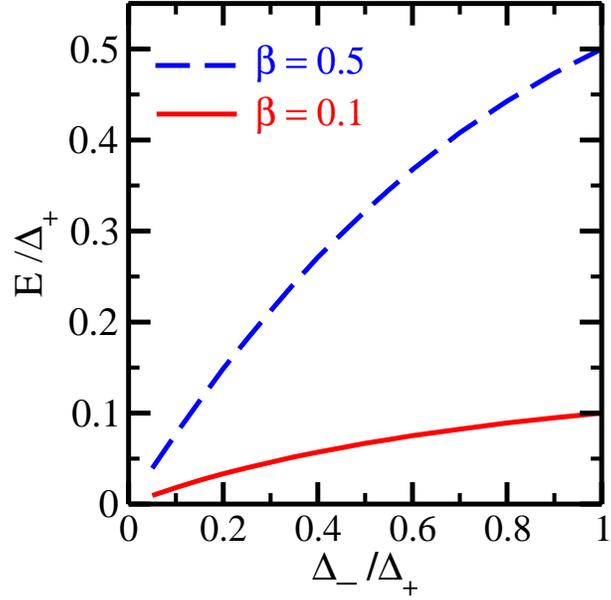}}
\caption{(Color online) The variation of edge state energy of the quasiparticles with the ratio of pair
amplitudes betwen two spin split bands for $\beta = 0.5$ and $0.1$. $E=\beta \Delta_+$ for $\Delta_-/\Delta_+ =1$
and $E$ converges towards zero for all values of $\Delta_-/\Delta_+$. However, $E=0$ only for $\beta =0$, i.e.,
$\phi_\pm =0$.}
\label{fig:energy}
\end{figure}

 The boundary condition $\Psi (x=0, y) =0$ determines the ratio between the coefficients $a$, $b$, $c$, and $d$
 and consequently we find an identity 
 \be
 \frac{ (\frac{u_+}{v_+})(\frac{u_-}{v_-}) +1}{\frac{u_+}{v_+}+\frac{u_-}{v_-}} = 
  \vert \beta \vert = \vert \frac{\sin [\frac{1}{2}(\phi_+ + \phi_-)]}
 {\cos [\frac{1}{2}(\phi_+ - \phi_-)]}\vert \, .
\label{identity}
\ee
Putting expressions of $u_+/v_+$ and $u_-/v_-$  in Eq.~(\ref{identity}), we find
\bea
 E^2+\Delta_+\Delta_- - \Gamma_+\Gamma_- -iE(\Gamma_+ + \Gamma_-) & &\nonumber \\
 =  \vert \beta \vert \left[ E(\Delta_-+\Delta_+) -i(\Delta_-\Gamma_+ +\Delta_+ \Gamma_-) \right] & & 
\label{identity2}
\eea
for positive energy quasiparticles.
An equivalent equation for edge state energy is also derived in Ref.~\onlinecite{Nagaosa}.
For a purely triplet superconductor, i.e., for $\Delta_+ = \Delta_-$, $E = \vert \beta\vert \Delta_+$. 
The solution of Eq.~(\ref{identity}) as a function
 $\gamma = \Delta_-/\Delta_+$ for $\beta = 0.5, 0.1$ is shown in Fig.~\ref{fig:energy}.
The zero energy edge state is possible only for $\beta = 0$ for all $\Delta_-/\Delta_+ > 0 \,\, ,\,\, (\Delta_- < \Delta_+)$.
There is no edge state for $\Delta_- =0$, i.e., when the triplet amplitude and singlet amplitude will
be of equal magnitude. This is because the superconductivity exists only in the
band of positive helicity as the negative helicity band becomes normal in this case. 
If $E= \Delta_-$, $u_- = v_-$ and consequently $\beta = \pm 1$ which suggests $ \vert \phi_+ \vert = \pi/2$.

When $\Delta_t < \Delta_s$, the pair amplitude in the negative helicity band is negative ($ \Delta_- <0 $). In that case
signs of third and fourth components of $\psi_e^-$ and $\psi_h^-$ in Eq.~(\ref{psi_minus}) change.  
Therefore Eq.~(\ref{identity2}) in this case reduces to
\bea
 E^2+\Delta_+\Delta_- - \Gamma_+\Gamma_- -iE(\Gamma_+ + \Gamma_-) & &\nonumber \\
 =  \vert\beta\vert \left[ -E( \Delta_+ +\Delta_-) +i(\Delta_- \Gamma_+ +\Delta_+ \Gamma_-) \right] & & \, .
\label{identity3}
\eea
This equation does not produce any solution in the range $\Delta_- \leq E \leq -\Delta_-$
except when the magnitudes of $\Delta_+$ and $\Delta_-$ are same and the corresponding
solution will be $E=\pm \Delta_+$. However, these solutions
do not correspond to edge state since $\kappa_\pm = 0$.
Therefore, there is no midgap edge bound state \cite{Nagaosa} for equal or larger singlet component compared to the triplet component.

\section{Charge and Spin Tunneling Conductance}

\begin{figure}
\centerline{\epsfysize=8cm\epsfbox{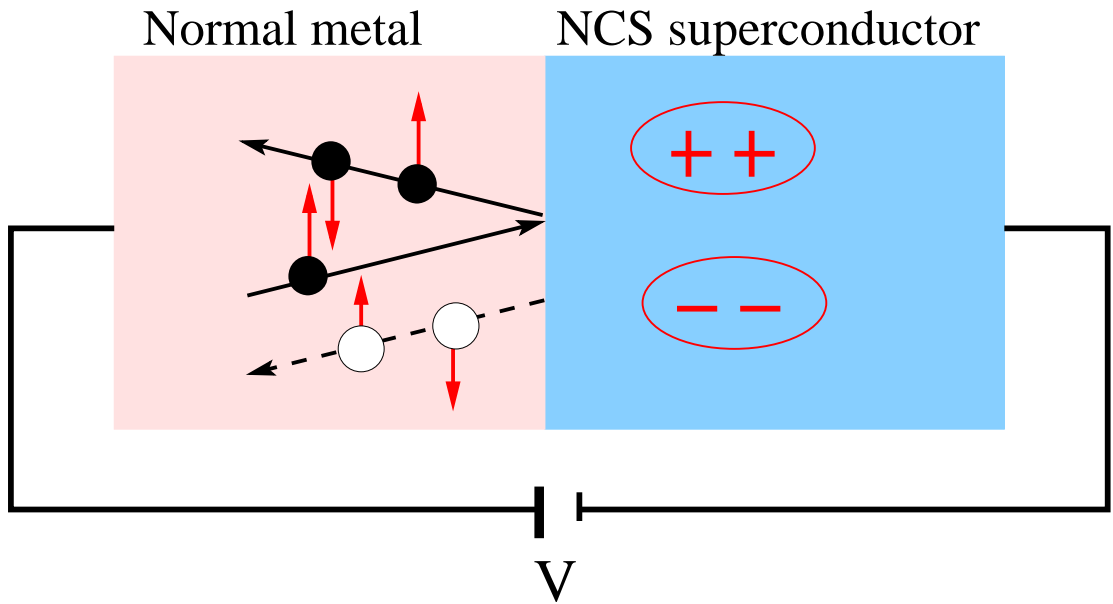}}
\caption{(Color online) A schematic diagram of tunneling from a normal metal to an NCS superconductor.
 Up (or down) spin electrons (filled circle) incident on the junction from the normal metal side gets partly reflected
as both spin-up and spin-down electrons as well as holes (open circle) in the Andreev process making
Cooper pairs inside the NCS superconductor at both positive and negative helicity bands. A bias voltage $V$
may be applied across the junction.}
\label{fig:cartoon}
\end{figure}

Consider a junction between a ballistic normal (at $x<0$) metal and an NCS (at $x>0$) superconductor. 
The junction is characterized by an insulating barrier at $x=0$ with a delta-function
potential $V(x) = U\delta(x)$. The Hamiltonian for the normal metal is $H_N = \xi_\k \,\hat{1}$.
In this geometry, the wave function for an electron with spin $\sigma \,( \text{numerically} \pm \, \text{ and symbolically}
\uparrow \text{or} \downarrow \text{respectively})$
incident from the normal metal on the junction is given by 
\bea
\Psi_N^\sigma (x,y) &=& e^{ik_yy}[(\psi_{e}^\sigma + a_{\sigma,\sigma} \psi^\sigma_h + a_{\sigma,-\sigma}\psi^{-\sigma}_h)e^{ik_{Fx}x} \nonumber \\
                  & & + (b_{\sigma,\sigma}+b_{\sigma,-\sigma}) \psi^\sigma_e e^{-ik_{Fx}x} ]
\eea 
within the "Andreev approximation",
where $^T\psi^\uparrow_e = (1,0,0,0),\, ^T\psi^\downarrow_e = (0,1,0,0),\, ^T\psi^\uparrow_h = (0,0,1,0),\, 
 ^T\psi^\downarrow_h = (0,0,0,1)$, and 
$k_{Fx} = \sqrt{k_F^2-k_y^2}$ with Fermi momentum $k_F$ in the normal metal. Here $a_{\sigma,\sigma}$, 
$a_{\sigma,-\sigma}$ $b_{\sigma,\sigma}$, and $b_{\sigma,-\sigma}$ are the
 parallel-spin Andreev, antiparallel-spin Andreev, parallel-spin normal, and antiparallel-spin normal
 reflection coefficients respectively. The normal and Andreev reflection processes and formation
of Coopar pairs inside the superconductor are schematically shown in  Fig.~\ref{fig:cartoon}.

The angle resolved charge and spin tunneling conductances are thus defined to be \cite{Blonder,Yoshida}
\bea
   g_c(\phi) &=& \left( 1 + \frac{1}{2} \sum_\sigma \left[ \vert a_{\sigma,\sigma} \vert^2  
                                 + \vert a_{\sigma,-\sigma} \vert^2 
                                 - \vert b_{\sigma,\sigma} \vert^2 
                                 - \vert b_{\sigma,-\sigma} \vert^2 
                \right] \right) \cos \phi \, , \label{GC1} \\
    g_s (\phi) &=& \left( \frac{1}{2} \sum_\sigma \sigma \left[ \vert a_{\sigma,\sigma} \vert^2
                                 - \vert a_{\sigma,-\sigma} \vert^2
                                 - \vert b_{\sigma,\sigma} \vert^2
                                 + \vert b_{\sigma,-\sigma} \vert^2
                \right]\right)   \cos \phi \label{GS1} 
\eea
respectively at zero temperature. Here the angle $\phi$ is defined as $k_y = k_F \sin\phi$. The reflection
amplitudes can be found out by matching the wave functions and the velocity flux at $x=0$:
\bea
& & \Psi_N^\sigma(x=0,y) = \Psi_S(x=0,y)  \,  , \label{BC1} \\
& & \left( \begin{array}{cccc}
-\frac{i}{m}\partial_x & 0 & 0 & 0 \\
0 & -\frac{i}{m}\partial_x & 0 & 0 \\
0 & 0 & \frac{i}{m}\partial_x & 0 \\
0 & 0 & 0 & \frac{i}{m}\partial_x 
\end{array} \right) \Psi_N^\sigma (x,y)\vert_{x=0} \nonumber \\ 
&=&  \left( \begin{array}{cccc} 
-\frac{i}{m}\partial_x & i\alpha & -i\frac{\Delta_t}{k_F} & 0 \\
-i\alpha & -\frac{i}{m}\partial_x & 0 & -i\frac{\Delta_t}{k_F} \\ 
i\frac{\Delta_t}{k_F} & 0 & \frac{i}{m}\partial_x & -i\alpha \\
0  & i\frac{\Delta_t}{k_F} & i\alpha & \frac{i}{m}\partial_x 
\end{array} \right) \Psi_S (x,y)\vert_{x=0} \nonumber \\
    & &   +2iU \left( \begin{array}{rrrr}
1 & 0 & 0 & 0 \\
0 & 1 & 0 & 0 \\
0 & 0 & -1 & 0 \\
0 & 0 & 0 & -1 
\end{array} \right) \Psi_N^\sigma(x=0,y) \,\, . \label{BC2}
\eea

\begin{figure}
\centerline{\epsfysize=9cm\epsfbox{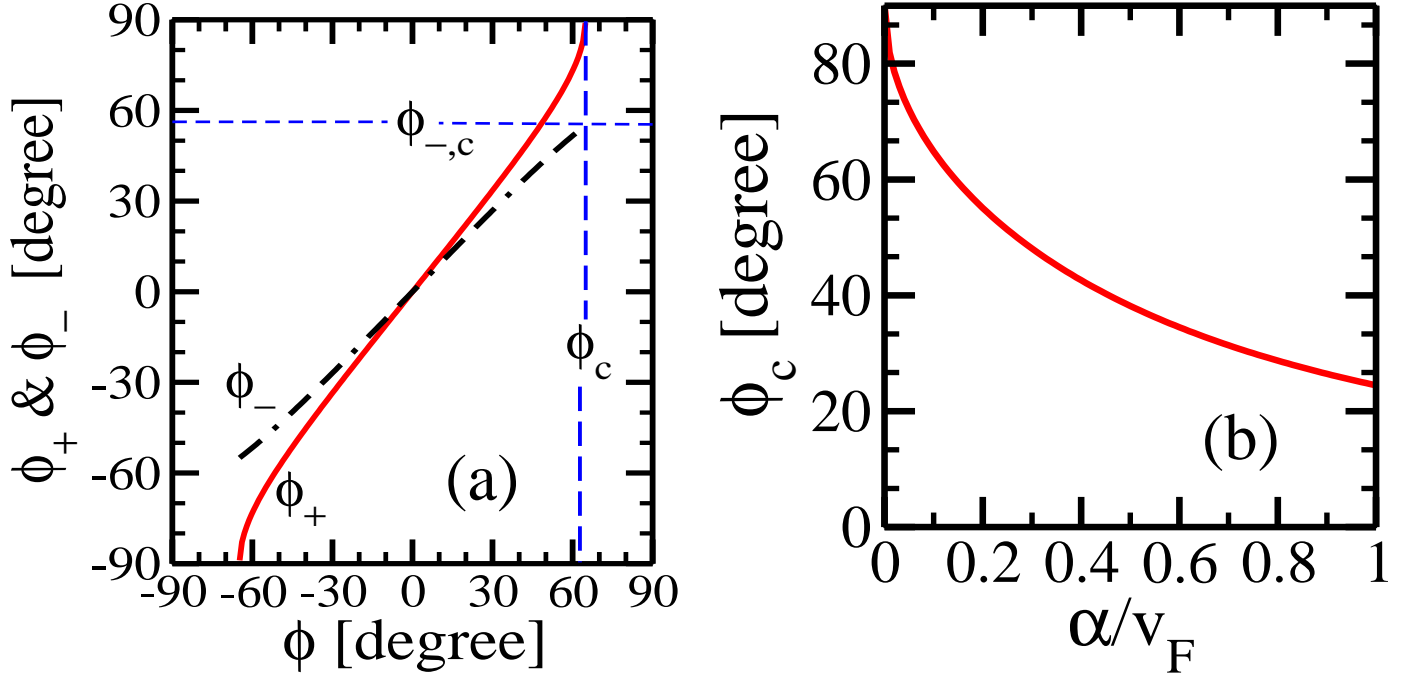}}
\caption{(Color online) (a) Variation of the angles $\phi_+$ and $\phi_-$  with the angle of incidence $\phi$ for
 $\alpha/v_F = 0.1$. The critical angle $\phi_c$ and correspondingly the critical angle
for negative helicity band, denoted as $\phi_{c,-}$ are shown. (b) Variation of $\phi_c$ against $\alpha /v_F$.}
\label{fig:critical_phi}
\end{figure}

For nonzero $\alpha$, the phase space of the incident electron that takes place in Andreev reflection
gets restricted. The angles of $\bm{k}$ in two bands inside the NCS superconductor is restricted
by $-\frac{\pi}{2} \leq \phi_\pm \leq \frac{\pi}{2}$. The conservation of momentum implies
$k_F\sin\phi = k_F^+\sin \phi_+ = k_F^-\sin\phi_-$. The variation of $\phi_\pm$ with the incident
angle $\phi$ is shown in Fig.~\ref{fig:critical_phi}(a) for $\alpha/v_F = 0.1$. It is clear that $-\phi_c \leq \phi \leq \phi_c$,
where $\phi_c$ is the critical angle of incidence beyond which incident electron becomes totally reflected.
This critical angle corresponds to $\phi_+ = \pi/2 $ and $\phi_- = \phi_{-,c}$.
The angle $\phi_c$ decreases with the increase of $\alpha$ as shown in Fig.~\ref{fig:critical_phi}(b). 
The total charge and spin tunneling conductances in the unit of normal tunneling charge conductance $G_{nc}$ become
\be
G_c = \frac{1}{G_{nc}}\int_{-\phi_c}^{\phi_c}g_c(\phi)\, d\phi\, ; \, 
G_s = \frac{1}{G_{nc}}\int_{-\phi_c}^{\phi_c}g_s(\phi)\, d\phi \, .
\label{GCS}
\ee

 We then consider the application of magnetic field $H$ perpendicular to the plane of the NCS superconductor.
Assuming the penetration depth is much larger than the coherence length of the superconductor, the corresponding
vector potential in the Landau gauge may be approximated as $\bm{A}(\bm{r}) = (0,-H\lambda_d \exp(-x/\lambda_d),0)$
with the penetration depth $\lambda_d$. In a semiclassical approximation where the quantization of the Landau level
may be neglected, the quasiparticle energy becomes Doppler shifted \cite{Sauls}: $E\to E-H\Delta_+\sin\phi /H_0$ with
characteristic filed $H_0 = \Phi_0/(\pi^2\xi\lambda_d)$, where coherence length $\xi = k_F/(\pi m\Delta_+)$ 
(as $\Delta_+$ is larger among two pair amplitudes) and
$\Phi_0$ is the flux quantum. The Zeeman coupling may be neglected since the energy of Doppler
shift energy is very high compared to the Zeeman energy for large $\lambda_d$. In contrast, the Zeeman energy is
responsible to break the degeneracy between the helical edge modes in quantum spin Hall systems (QSHS) \cite{Kane,Zhang,Molenkamp} and
it modulates the transport properties.
The modulation of the spin conductance with H due to the Doppler shift in NCS superconductor
as superconducting analogue \cite{Nagaosa} to the QSHS as topological system.

In presence of small magnetic field, where the formation of Landau levels are ignored, the wave function in
the normal side remain as superposition of plane waves as in the case of zero magnetic field.
We also ignore the spin reflection asymmetry arising from Zeeman coupling in the normal side.
We numerically evaluate the coefficients $a$'s and $b$'s, both in the absence and presence of magnetic field,
using Eqs.(\ref{BC1}) and (\ref{BC2}) and plug them into
Eqs.(\ref{GC1}) and (\ref{GS1}) to determine angle resolved charge and spin conductances. The total charge and spin
conductances are then evaluated using Eq.(\ref{GCS}). The numerical results are presented below for a fixed parameter
$Z=2U/v_F $ characterizing the effective strength of the barrier. However, the qualitative behavior is independent
of $Z$ as we see below.

\section{Results}

\begin{figure}
\centerline{\epsfysize=8cm\epsfbox{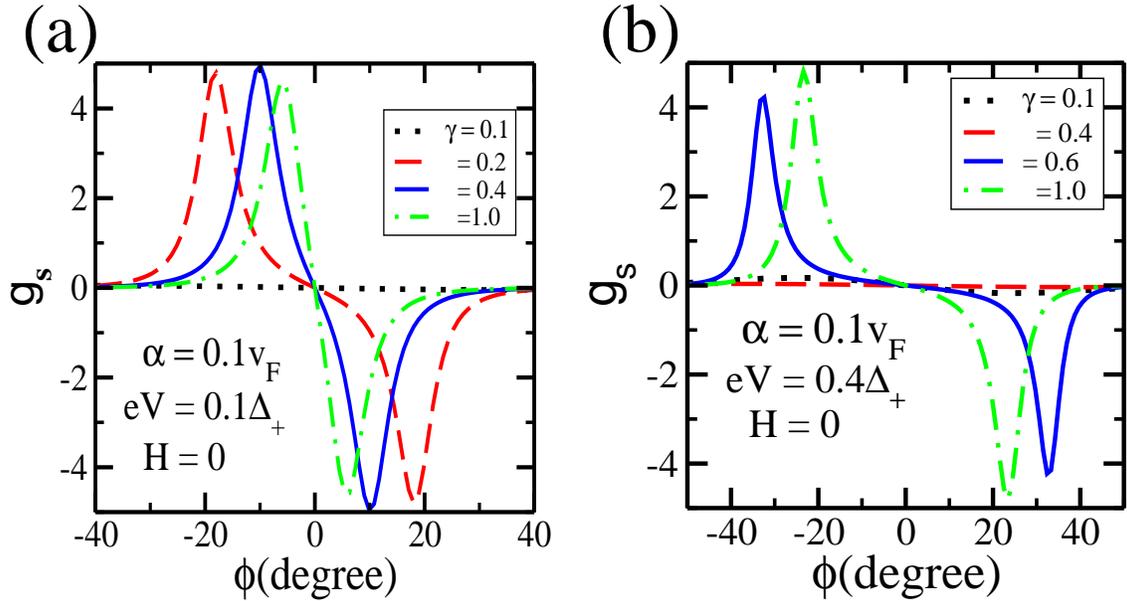}}
\caption{(Color online) The variation of spin conductance $g_s$ (in the unit of $G_{nc}$) with the incident angle $\phi$ for different values of $\gamma$
at quasiparticle energy $E=eV=0.1\Delta_+$ (a) and $0.4\Delta_+$ (b) for $H=0$, $Z=5$, and $\alpha = 0.1v_F$.}
\label{fig:spincond_phi}
\end{figure}

Although the NCS superconductors do not break time reversal symmetry, angle resolved spin conductance is nonvanishing and
$g_s(\phi)$ shows peaks at those values of $\phi$ for which energy of incident electron matches with the
energy of the midgap edge state. The large $g_s(\phi)$ is due to the presence of helical edge modes \cite{Nagaosa} in
NCS superconductors. 
We have found that $g_s(\phi)$ depends very weakly on $\alpha/v_F$ around the peak position. 
The variation of $g_s(\phi)$ is shown in Fig.~\ref{fig:spincond_phi} for different values of the ratio $\gamma=\Delta_-/\Delta_+$,
and two different values of quasiparticle energy $eV$ for a bias voltage $V$ across the junction.
The peak in $g_s(\phi)$ is present for $\vert eV\vert < \Delta_-$. The peak shifts towards smaller $\vert \phi \vert$
for larger values of $\gamma$.
However, the total spin conductance becomes zero since $g_s(\phi) = -g_s(-\phi)$ for any values of $\alpha/v_F$, $\gamma$, and $eV$.

\begin{figure}
\centerline{\epsfysize=20cm\epsfbox{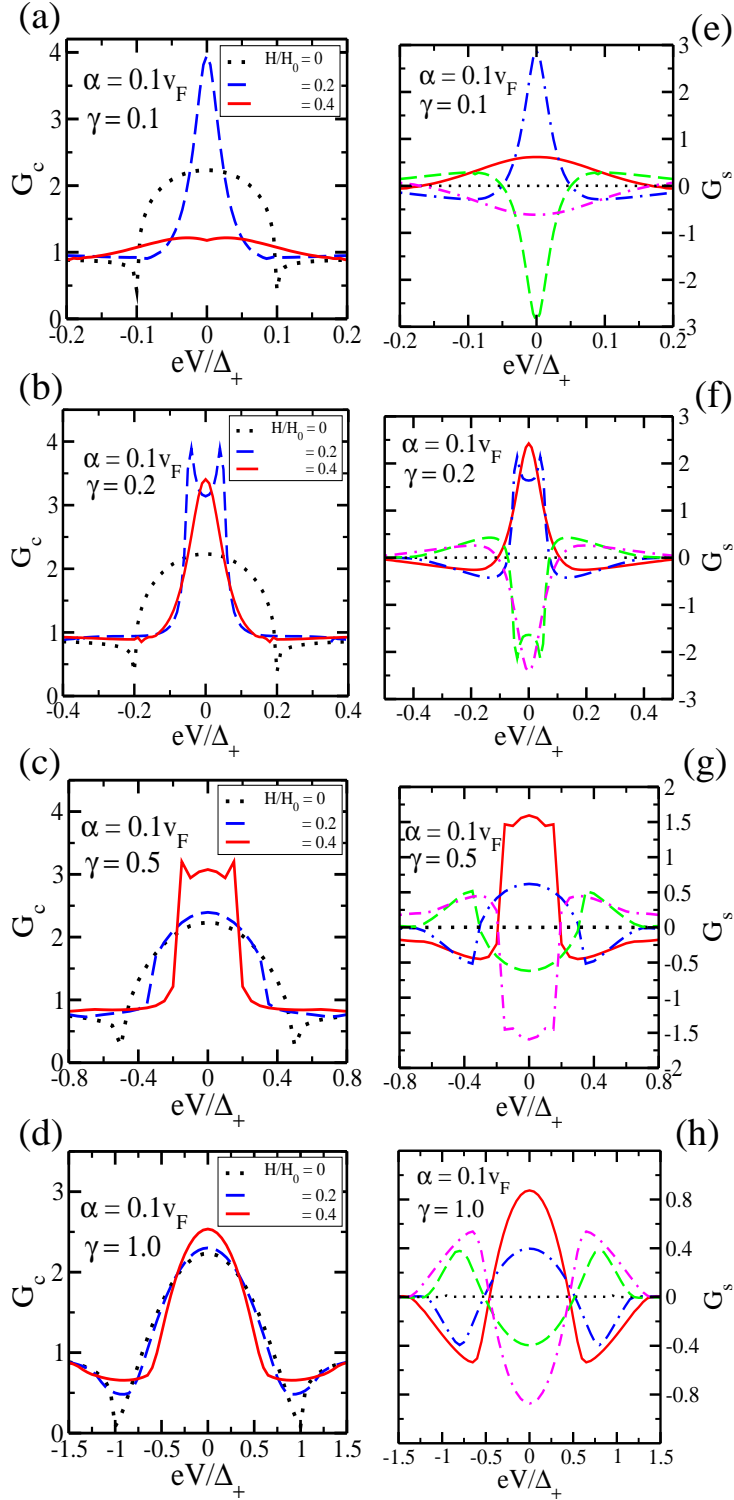}}
\caption{(Color online) The variation of charge conductance (a--d) and spin conductance (e--h) with bias energy $eV$.
The parameters $Z=5$, $\gamma = 0.1$ (a,e), 0.2 (b,f), 0.5 (c,g), and 1.0 (d,h), and $\alpha/v_F = 0.1$ are chosen.
The magnetic filed $H/H_0$ chosen for the panels (e--h) are -0.4 (solid line), -0.2 (dot and long-dashed line, 0
(dotted line), 0.2 (dashed line), and 0.4 (dot and short-dashed line).
}
\label{fig:cond_mag}
\end{figure}

The total charge and spin conductances for different values of $H$ and $\gamma$ are shown in Fig.~\ref{fig:cond_mag}. 
Since $G_c$ and $G_s$
are weakly dependent on $\alpha$, we choose a fixed value $\alpha/v_F = 0.1$. The charge conductance is minimum
at $\vert eV\vert = \Delta_-$ in the absence of magnetic field since $\Delta_-$ is the lowest energy scale 
in the bulk superconductor. 
The zero bias peak in $G_c$
at $H=0$ is present as is observed \cite{Covington,Sauls,Aprili} in $d$-wave and predicted \cite{Tanaka2,Tanaka3} 
in $p$-wave superconductors. 
When the bound state quasiparticle energy $E=\Delta_-$, $\phi = \phi_c$. In that case $\gamma = \vert eV 
\pm \frac{H}{H_0}\sin \phi_c \vert$ in presence of bias and magnetic field.  
The zero bias peak remains for $\frac{\vert H\vert }{H_0}> \frac{\gamma}{\sin \phi_c}$ but $G_c$ decreases with the
increase of $H$ at high magnetic field. The ZBP in $G_c$ at finite manetic field splits into two sharp
peaks at finite bias
(one at negative bias and the other at positive bias) and a dip in zero bias,
when $\gamma > \frac{\vert H\vert }{H_0}\sin \phi_c$. 
The peaks shift towards higher $\vert eV\vert$ and becomes weaker on lowering $\vert H\vert $ so that
the ZBP reappears again at a low field. 
Although the total spin conductance $G_s$ is zero at any bias, it modulates with $eV$ at finite $H$. It has the
symmetry: $G_s(eV,H) = -G_s(eV,-H)=-G_s(-eV,H)$. The disappearance and reappearance of ZBP in 
the magnitude of $G_s$ and the splitting of ZBP at finite magnetic field is similar to that of $G_c$.

\begin{figure}
\centerline{\epsfysize=8cm\epsfbox{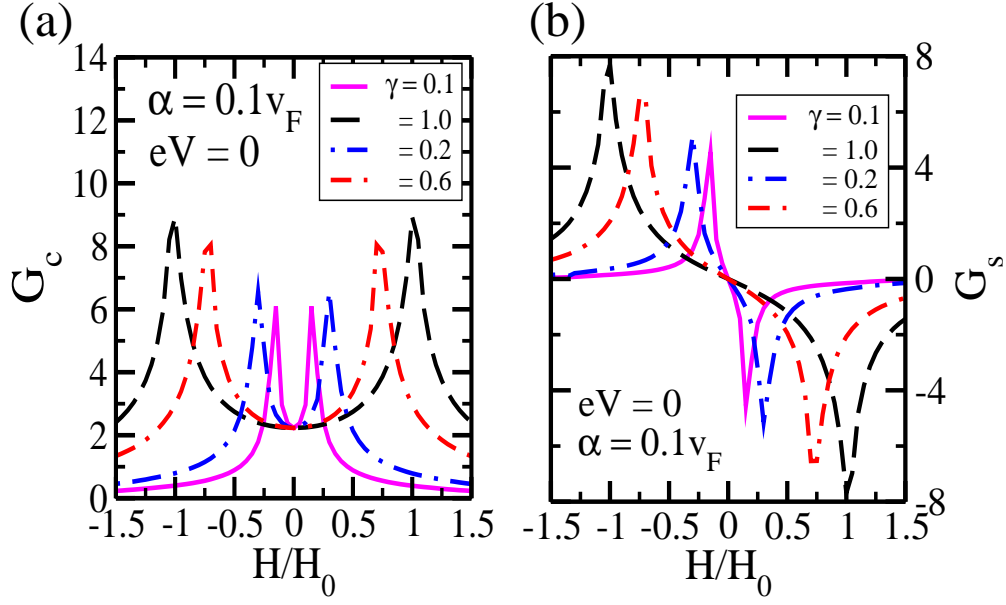}}
\caption{(Color online) (a) Charge conductance and (b) spin conductance {\it vs}. $H/H_0$ for $Z= 5$, $\alpha/V_F = 0.1$ and at zero
bias. The curves from left correspond to $\gamma = 1.0$, 0.6, 0.2, and 0.1.}
\label{fig:cond_zero_bias}
\end{figure}

The ZBP in $G_c$ increases initially with the magnetic field and it subsequently decreases creating
a peak at $\vert H\vert \sim \gamma H_0$, i.e., when all the midgap edge states upto the energy $\Delta_-$
take part in the conduction process. Likewise ZBP in $G_s$ also behave sameway with the important exception
that the latter changes sign on reversing the magnetic filed direction,
although $G_s$ is zero at $H=0$. This is an extraordinary effect on the spin as well as charge conductances as shown
in Fig.~\ref{fig:cond_zero_bias}
 by the presence of midgap helical edge states. We observe that the value of $\vert H\vert /H_0$ at which the peaks
occur decreases with $\gamma$ since $\Delta_-$ decreases with a fixed $\Delta_+$.    
For $\gamma =1$, the ZBP in $G_c$ is almost constant at small $H \, (\vert H \vert < 0.5H_0)$ but the ZBP in
 $G_s$ changes sharply at small $H$ as is obtained by
Tanaka et al \cite{Nagaosa}. However when $\gamma$ is small, the ZBP in both $G_c$ and $G_s$ form peaks at much smaller field.
In the system like Li$_2$Pt$_3$B (Ref.\onlinecite{Yuan}), spin-triplet and -singlet components are in same order which
means $\gamma$ is small and it is estimated to be $\sim 0.24$. 
Therefore in such systems the presence of helical edge states will be revealed in form of peaks for
zero bias charge and spin magneto-tunneling-conductance at as small as $\sim 0.35H_0 \sim
0.07$ Tesla  magnetic field for typical values of $\xi \sim 10$ nm and $\lambda_d \sim 100$ nm.

\begin{figure}
\centerline{\epsfysize=4cm\epsfbox{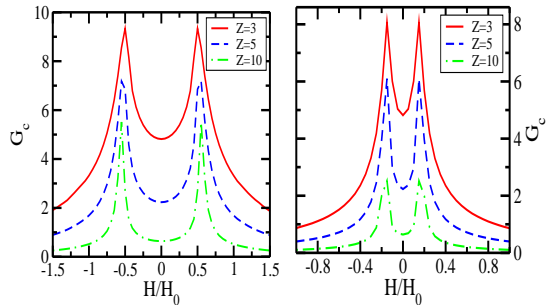}}
\caption{(Color online)  Variation of zero bias charge conductance with $H$ at barrier heights $Z= 3, 5$, and 10 
with $\gamma = 0.4$ (left panel) and 0.1 (right panel)
when $\alpha/V_F = 0.1$}. 
\label{fig:zvariation}
\end{figure}

 In our study so far, we have chosen $Z=5$ as the parameter for barrier height. Fig.~\ref{fig:zvariation} shows
the variation of $G_c$ at zero bias as a function of magnetic field for different values of $Z$ and $\gamma$.
We notice that the qualitative behaviour, in particular the positions of ZBP are independent of $Z$. The values of 
the tunneling conductances increase with decreasing $Z$, as expected.

\section{Summary}

To summarize, the helical edge states \cite{Nagaosa} exist in a noncentrosymmetric superconductor provided the
triplet-pair-amplitude is larger than the singlet-pair-amplitude, i.e., when $0 < \gamma \leq 1$. The energies of the 
midgap ($E<\Delta_-$) edge states decrease with $\gamma$. We have studied the consequence of these edge states on
the charge and spin tunneling conductances from a normal metal to a noncentrosymmetric superconductor.   
The angle resolved spin conductance $g_s$ shows peak at an angle that correspond to the conduction through the edge state.
The $g_s$ show peaks when the bias energy $\vert eV \vert < \Delta_-$. 
It changes sign on the reversal of sign of the angle since the conduction is due to helical edge states and
this change of sign leads to zero total spin conductane $G_s$ irrespective of the bias. However, the Doppler shifted
energy of the quasiparticles for the application of $H$ leads to nonzero $G_s$ and it modulates with $eV$ for different magnetic
fields. The zero bias peak is present at high $H$ (although $G_s$ vanishes at very high $H$). This peak splits into two
(one at positive bias and the other at negative bias) and a dip is formed at zero bias on reduction of the field. The double
peaks occur when $\gamma > \frac{\vert H \vert}{H_0}\sin \phi_c$ and they become weaker on lowering the field so that a zero
bias peak reappear at very low field. Similarly, the disappearance and reappearance of zero bias peak in total charge
conductance $G_c$ also occur. Moreover, $G_c$ has a dip at $\vert eV\vert = \Delta_-$. Interestingly, the magnitude of
zero bias charge and spin magneto-conductance increases with $\vert H\vert$ and form peaks at $\vert H\vert \sim \gamma H_0$,
i.e., when all the midgap helical edge states take part in the conduction process.

\section*{Acknowledgment}
SPM is supported by CSIR, Government of India.

\newpage

\newpage

\newpage

\newpage

\newpage

\newpage

\end{document}